\documentclass[structabstract]{aa}  
\usepackage{graphicx}
\usepackage{txfonts}
\usepackage{footmisc}
\usepackage{float}

\providecommand{\abs}[1]{\left|{#1}\right|}
\long\def\symbolfootnote[#1]#2{\begingroup\def\thefootnote{\fnsymbol{footnote}}
\footnote[#1]{#2}\endgroup}
\renewcommand{\thefootnote}{\fnsymbol{footnote}}
\begin{document}
%
   \title{Conversion from linear to circular polarization in FPGA }

   \author{Koyel Das
          \inst{1},
	   A. L. Roy
	  \inst{1},
	  R. Keller
	  \inst{1}	
          \and
          G. Tuccari\inst{2}
          }

   \institute{Max Planck Institute for Radio Astronomy (MPIFR), 
              Auf dem H\"ugel 69, Bonn, Germany\\
              \email{koyel@mpifr-bonn.mpg.de, aroy@mpifr-bonn.mpg.de, rkeller@mpifr-bonn.mpg.de}
         \and
             Istituto di Radioastronomia, via P. Gobetti, 101 40129 Bologna, Italy\\
             \email{g.tuccari@ira.inaf.it}
             }


 
  \abstract
    {Radio astronomical receivers are
     now expanding their frequency range to cover large (octave) fractional
     bandwidths for sensitivity and spectral flexibility, which makes the
     design of good analogue circular polarizers challenging. Better polarization purity requires a flatter phase response over increasingly wide bandwidth, which is most easily achieved with digital techniques.
     They offer the ability to form circular polarization with perfect 
     polarization purity over arbitrarily wide fractional bandwidths, due
     to the ease of introducing a perfect quadrature phase shift.
     Further,
     the rapid improvements in field programmable gate arrays provide 
     the high processing power, low cost, portability and reconfigurability
     needed to make practical the implementation of the formation of 
     circular polarization digitally.}
   {Here we explore the performance of a circular polarizer implemented with digital techniques.}
   {We designed a digital circular polarizer in which the intermediate frequency signals from a receiver 
with native linear polarizations were sampled and converted to circular
polarization. The frequency-dependent instrumental phase difference and gain scaling factors were determined using an injected noise signal and applied to the two linear polarizations to equalize the transfer characteristics of the two polarization channels.  This equalization was performed in 512 frequency channels over a 512 MHz bandwidth.  Circular polarization was formed
by quadrature phase shifting and summing the equalized linear polarization signals.}
   {We obtained polarization purity of -25 dB corresponding to a D-term of 0.06 over the whole bandwidth.}
   {This technique enables construction of broad-band radio astronomy receivers with native 
linear polarization to form circular polarization for VLBI.}

   \keywords{formation of circular polarization -- instrumental phase and gain calibration -- real time application
               }

   \maketitle

%
%

\section{introduction}

Circular polarizers play important roles in modern communication systems including those in radio astronomy. To obtain higher sensitivity and frequency coverage for spectral line observations, the radio antennas are moving to broad-band feeds and extremely broad bands are most easily realized with linearly polarized feeds due to the difficulty of producing $90^{\circ}$  phase shift accurately over wide bandwidth. However, circular polarization is simplest for the application of very long baseline interferometry (VLBI), which enables astronomical sources to be resolved with sub-milliarcsecond synthesized beam widths, since linear dipoles do not generally remain
parallel to each other in a global array due to different parallactic
angles at different stations when observing the same source, causing
loss of coherence in the cross-correlation products formed between
stations.  That loss could be recovered were one to compute also the cross-polarization cross
correlation products to retain all information (doubling the
correlator power needed), or one could rotate the receiver packages at
each station to keep the dipoles parallel (requiring mechanical rotators).  In contrast, use of
circular polarization causes the parallactic angle differences between
stations to add a simple phase rotation angle to the measured visibility,
which can be predicted from the known observation geometry and
subtracted in post-processing.\\
\\
Circular polarizers with broad bandwidths have been realized in the
past with a number of methods.  Most common are the septum polarizer
by \cite{chen73} , corrugated waveguide phase shifter, Boifot junction \cite{ruiz06}, and
the linear quad-ridge OMT followed by a $90^{\circ}$ hybrid junction.
All are analogue techniques and produce a perfect $90^{\circ}$ phase
shift and hence perfect polarization purity at only one, two, or three
frequencies and the phase errors grow larger at frequencies away
from those design points, which ultimately limits the bandwidth of the 
devices. In contrast, digital techniques offer the possibility to produce an
accurate $90^{\circ}$ phase shift over broad bandwidths, but this
potential has not yet been fully developed.
One example is the Westerbork synthesis radio telescope, which
converts native linear polarization to circular polarization by a combination of
analogue and digital techniques for VLBI.  During down conversion of
the orthogonal linear signals, the $90^{\circ}$ shift is added to the
(analogue) LO for one polarization.  After analogue-to-digital
conversion, the (2 bit) signals are summed and differenced to form
circular polarization, with a weight that corrects for the average
receiver gain differences.  The weights are determined by a separate
measurement using a calibration noise source in each frontend and are
updated every 10 s.  This system operates on a bandwidth of 
20 MHz, yielding one phase and amplitude correction for each 20 MHz of
bandwidth (Boss 2007, private communication). 
The Westerbork system uses the pre-existing correlator and analogue
phase rotation in the LO system, but most VLBI stations lack this
equipment and so another, more general, solution is needed.\\
\\
We have developed a self-contained digital processing system in which
the correlator, channel equalization, phase rotation, gain
scaling, quadrature phase shift and summation to form circular polarization are contained in a stand-alone unit.  It accepts two intermediate frequency inputs
with orthogonal linearly polarized signals, each of 500 MHz bandwidth
and subdivides the band down to 1 MHz resolution.  In each 1 MHz piece it
measures the phase and amplitude differences between the orthogonally
polarized channels using a calibration noise source in the front
end that is common to both polarizations.  It then uses those measurements
to equalize the channel phases and amplitudes during observations.
After equalization it introduces an ideal $90^{\circ}$ phase shift into
one polarization channel and forms sum and difference outputs that respond to orthogonal
circular polarizations at the input. \\
\\
Digital systems offer the ability to process continuous data in real time implementing automatic data processing algorithms. Data transmitted digitally are more resistant to external interference and hence digital devices supersede analogue counterparts at least where speed and signal purity are matters of concern. With the advent of logic devices like  ASICs and FPGAs, it is possible to implement many complex algorithms, which would have been impossible otherwise and they also offer the user ease of replication compared to analogue systems. For these reasons our system to produce circular polarization generates almost ideally perfect orthogonal field components in real time.\\
\\
This paper acquaints the reader with the theory and derivation for obtaining instrumental phase and gain correction factors and applying them to the two received orthogonal linear polarizations to form phase and gain calibrated left hand circular (LHC) and right hand circular (RHC) polarizations. It explores limitations of system performance due to the most influential factors, which are the D-terms and receiver instabilities and explores the requirement for periodic recalibration to remove their detrimental effects on polarization purity. It describes the digital signal processing in FPGA and the experimental verification of the technique to show that good polarization purity is obtained.


\section{Theoretical Development}
2.1 Overview of Method:
We measure the receiver transfer characteristics using a noise diode signal as
a calibration source, injected into the feed as pure linear polarization in a plane oriented at $45^{\circ}$ to the two orthogonal linear dipoles that couple energy out of the waveguide.  The noise diode injects signals into the $x$ and $y$ chains that are identical, having equal amplitude, zero phase difference, and 100 \% 
coherence with each other.  These noise diode signals during
calibration pass through the same $x$ and $y$ receiving chains as does the
astronomical signal later, and are sampled at IF at 1 GS/s (i.e. 500 MHz
contiguous Nyquist bandwidth).  The sampled $x(t)$ and $y(t)$ signals are
processed in an FX correlator on the FPGA, which transforms to
frequency domain with 1 MHz channel widths, cross multiplies each $X(\omega)$
spectrum against the corresponding $Y(\omega)$ spectrum and integrates for 8 s. We chose a spectral resolution of 1 MHz to allow for possibly rapidly changing channel phase differences with frequency. The result is a phase spectrum with low thermal phase noise that represents the phase difference between the $x$ and $y$ channels due to
the transfer characteristics of the receiver chains.  The phase
spectrum is used during later astronomical observation for equalizing
the (frequency-dependent) phase lengths of the $x$ and $y$ receiver
chains.  The $x$ and $y$ bandpass amplitude shapes are also equalized,
using gains derived during calibration from total-power spectra of $X(\omega)$
and $Y(\omega)$ accumulated during the calibration stage.\\
To form circular polarization from native linears during astronomical
observations, we need likewise to transform the $x(t)$ and $y(t)$ time series
to frequency domain in the same manner as during calibration, then
equalize the transfer characteristics by applying a phase rotation 
to each frequency channel of one polarization and an amplitude scaling 
to each frequency channel of both polarizations in an equalizer stage,
then simply add or subtract $90^{\circ}$ (equivalent to exchanging real
and imaginary in the complex spectra), and summing to form results
that respond to the two hands of circular polarization.\\
\\
2.2  Instrumental Phase and Gain Calibration: 
The orthogonal time-domain field components $x(t)$ and $y(t)$ received by the crossed dipoles undergo unequal phase and amplitude distortions due to different frequency-dependent time delays and gains of the two receiving systems through which they pass.  The phase and gain calibration aims to compensate  the transfer characteristics of channel $x$ and channel $y$ to make them identical in $x(t)$ and $y(t)$, reducing instrumental artifacts to zero. This approach is already used in software for calibrating radio astronomical data though those operate on stored data rather than in real time. Our effort is to extend it to the digital domain processing sampled IF signals in real time, calibrating with fine frequency channels, to enable formation of circular polarization in real time with more accurate phase and magnitude response than the analogue techniques can achieve over broad bandwidths. We have not considered channel non-linearities and multi-path effects. Non-linearity spreads the output spectrum beyond the input spectrum by introducing new frequency components and causes amplitude distortion. Therefore, radio astronomical receivers have to be designed to be linear. In the presence of strong RFI non-linearity does occur and has to be blanked. Techniques for RFI mitigation are a sizable study in themselves and are beyond the scope of the present work. Nevertheless RFI mitigation techniques can easily be implemented in the same digital hardware. For the present development we assume linear transfer characteristics and Gaussian signal statistics. We have filtered the passband to avoid aliasing and kept signal levels in the linear regime. A linear  time invariant system causes only pulse dispersion and amplitude scaling.\\
\\
2.3 Phase Equalization:
Let us consider the noise diode signal during calibration as  a broadband source radiating Gaussian random signals continuously in time, $s(t)$, and we receive and sample a finite number, $N_s$, of frames of time-domain samples each consisting of $N$ samples spaced equally in time in two orthogonal linear polarization states, $x(t)$ and $y(t)$. Let the sampled time series be represented by $x_i(t)$ and $y_i(t)$ and the noise diode signal at the sample times be $s_i(t)$ where $i = 1,2,3,\ldots,N_s$. It is convenient to transform these time series into the frequency domain since the transfer characteristic of the receiving system used for calibration is frequency dependent. The Fourier transform produces $N_s$ spectra, each consisting of $N$ channels.  $N_s$ depends on the sampling rate, $f_s$, the number of samples in a spectrum, $N$, and the total integration time, $T_{integ}$, as
 	\begin{equation}
	N_s = \frac{f_sT_{integ}}{N}.
	\end{equation}
Let a time-domain signal have length $T_0$, then the corresponding frequency-domain spectrum will have channels spaced at an interval $f_0$ of $\frac{1}{T_0}$ frequency units. For a simplified analysis we  consider one frequency component, the results from which hold good for all other spectral components in the band.
The signals $x_i(t)$ and $y_i(t)$ are represented by the equations
 	\begin{equation}\label{eqn:2.1}
	x_i(t) = h_x(t) * (s_i(t) + n_{xi}(t)), \rm{and}
	\end{equation}
	\begin{equation}\label{eqn:2.2}
	y_i(t) = h_y(t) * (s_i(t - t_{xy}) + n_{yi}(t))
	\end{equation}
where $h_x(t)$ and $h_y(t)$ are the transfer functions of channel  $x$ and channel $y$, $t_{xy}$ is the $x-y$ propagation time difference from the dipoles to the receiver inputs (samplers for a digital receiver), and $n_{xi}(t)$ and $n_{yi}(t)$ are external unwanted signals (astronomical sources, thermal fluctuations and spurious sources). After transforming to the frequency domain these relations are 
	\begin{eqnarray}\label{eqn:2.3}
X_i(\omega)&=&\abs{X_i(\omega)}  e^{j  \phi_X(\omega)}\nonumber\\
&=&\abs{H_X(\omega)}  e^{j  \theta_X(\omega)}  (\abs{S_i(\omega)} + N_{Xi}(\omega)), \rm{and}
	\end{eqnarray}
	\begin{eqnarray}\label{eqn:2.4}
Y_i(\omega)&=&\abs{Y_i(\omega)}  e^{j  \phi_Y(\omega)}\nonumber\\
&=&\abs{H_Y(\omega)}  e^{j  \theta_Y(\omega)}  (\abs{S_i(\omega)}  e^{-j  \omega  t_{xy}} + N_{Yi}(\omega))
	\end{eqnarray}
where $\phi_X(\omega)$ and $\phi_Y(\omega)$ are the phases of $X_i(\omega)$ and $Y_i(\omega)$ respectively and $\theta_X(\omega)$ and $\theta_Y(\omega)$ are the phases of the transfer functions $H_X(\omega)$ and $H_Y(\omega)$ respectively. The phase difference $\phi_X(\omega) - \phi_Y(\omega)$ is due to the initial and instrumental phase difference between $X(\omega)$ and $Y(\omega)$. 
Now, let us consider the samples of $X_i(\omega)$ and $Y_i(\omega)$ at uniform intervals of $\omega_0=2\pi f_0$. If $X_i(r\omega_0)$ and $Y_i(r\omega_0)$ are the $r^{th}$ (channel number) samples of $X_i(\omega)$ and $Y_i(\omega)$ respectively, where $r= 0,1,2,3,\ldots,N-1$, then equations $(\ref{eqn:2.3})$ and $(\ref{eqn:2.4})$ can be rewritten in discrete form \cite{cooley65} as
	\begin{eqnarray}\label{eqn:2.5}
           X_i(r\omega_0)&=&\abs{X_i(r\omega_0)}  e^{j\phi_X(r\omega_0)}\nonumber\\
&=&\abs{H_X(r\omega_0)}  e^{j\theta_X(r\omega_0)}\nonumber\\
&&(\abs{S_i(r\omega_0)} + N_{Xi}(r\omega_0))
	\end{eqnarray}
	\begin{eqnarray}\label{eqn:2.6}
           Y_i(r\omega_0)&=&\abs{Y_i(r\omega_0)}  e^{j\phi_Y(r\omega_0)}\nonumber\\
&=&\abs{H_Y(r\omega_0)}  e^{j\theta_Y(r\omega_0)}\nonumber\\
&&(\abs{S_i(r\omega_0)}  e^{-j  r\omega_0  K_1T_s} + N_{Yi}(r\omega_0))
	\end{eqnarray}
	where $K_1$ is the fractional sample delay caused by $t_{xy}$ and $T_s$ is the sampling period. It is assumed that the system is linear. Thus the transfer function causes only linear distortion and no new frequency components are produced. The resulting pulse is dispersed in time and amplitude rescaled.
	The phase difference $\phi_X(r\omega_0) - \phi_Y(r\omega_0)$ is obtained from the accumulated 	cross power spectrum of $X_i(r\omega_0)$ and $Y_i(r\omega_0)$ in the frequency domain, 	which is expressed as
	\begin{eqnarray}\label{eqn:2.7}
Z(r\omega_0) &=& \sum_{i=1}^{N_s} Z_i(r\omega_0) \nonumber\\
&=& \sum_{i=1}^{N_s} \abs{X_i(r\omega_0)}  \abs{Y_i(r\omega_0)}  e^{j(\phi_X(r\omega_0) - \phi_Y(r\omega_0))}.
	\end{eqnarray}
	Substituting Eq.(6) and Eq.(7) in Eq.(8) we obtain the product $\abs{X_i(r\omega_0)}  \abs{Y_i(r\omega_0)}  e^{j(\phi_X(r\omega_0) - \phi_Y(r\omega_0))}$, which consists of the summation of the following contributing terms. \begin{eqnarray}\label{eqn:2.8}
1.&\abs{H(r\omega_0)}&  e^{j\theta(r\omega_0)}  \abs{S_i(r\omega_0)}^2  e^{jr\omega_0  K_1T_s}\\
2.&\abs{H(r\omega_0)}&  e^{j\theta(r\omega_0)}  N_{Yi}^{*}(r\omega_0)  N_{Xi}(r\omega_0)
	\end{eqnarray}
where $\abs{H(r\omega_0)}  e^{j\theta(r\omega_0)}=\abs{H_X(r\omega_0)}  \abs{H_Y(r\omega_0)}  e^{j(\theta_X(r\omega_0) - \theta_Y(r\omega_0))} $.
$Z_i(r\omega_0)$ is accumulated for $T_{integ}$ time. Since $S_i$ is incoherent with $N_{Xi}$ and $N_{Yi}$, their cross product terms will average to zero in the summation and so are not shown here. We have chosen, in practice, 8 s integration time corresponding to $8\times10^6$ frames of data so that the thermal fluctuations of the noise diode signal are averaged down to a fractional fluctuation of $4\times10^{-4}$ in 1 MHz channels that is 5 \% of the system temperature. Thus we attain our objective of measuring the transfer characteristic phase to  $\ll0.1^{\circ}$ precision. At such high levels of precision, the measurement is sensitive to corruption by possible external sources or by RFI via the second term Eq.$(10)$. We cancel this effect on our measurement of the transfer characteristic, provided the corrupting source remains constant, by performing the summation in Eq.$(8)$ twice, first in the presence of $s_i(t)$ (noise diode switched on) and second with $s_i(t) = 0$ (noise diode switched off) and differencing. Hence, considering only the first term (Eq.$9$) we obtain the phase difference $\phi$ between the two polarization signals due to instrumental effects and due to any initial phase difference as follows. 
	\begin{eqnarray}\label{eqn:2.13}
	\phi = \phi_{instrument} + \phi_{initial}
	\end{eqnarray}
	where, $\phi_{instrument} = \theta(r\omega_0)$ and $\phi_{initial} = r\omega_{0}  K_1T_s$ \footnote[3]{We have not considered any frequency down conversion in Eqs. (6) and (7). A down conversion will cause a frequency shift in $\omega$ by an amount $\omega_d$ where $\omega_d$ is the  mixing frequency.  Then the samples of $\omega$ in Eqs. (6) and (7) or $r\omega_0$ will represent the samples of  $\omega -\omega_d$. The equations will remain unchanged except for the initial phase, which will now be $(r\omega_0 + \omega_d)K_1T_s$ so that the initial phase is unaffected by down conversion. }. 
	For accurate calibration, $\phi_{instrument}$ should be measured and used later to correct the signals for the effect of $\phi_{instrument}$. This should reduce the phase difference to zero, requiring that the initial phase difference be zero so that $\phi$ corresponds to instrumental phase errors only. 
	This requires that the two orthogonal dipoles receive signal from a common source placed at a location equidistant from both the dipoles, which is accomplished by placing a noise source at $45^{\circ}$ to the two dipoles. One could alternatively calibrate using an unpolarized astronomical source instead of the noise diode, and point on source and then off source for the two accumulations to be differenced. In the on-source position, the dish main beam must be pointed accurately to the source, placing the source on the focal axis. The dipoles are located in a plane accurately perpendicular to the focal axis and hence, the dipoles would be equidistant 	from the source to high accuracy, so $\phi_{initial}$ would be close to zero and can be ignored. $Z(r\omega_0)$ gives the angle through which one of the vectors say $Y(r\omega_0)$ must be rotated to make the phase lengths of the two polarization channels equal. The frequency-dependent rotation matrix elements $\rm{cos}\theta(r\omega_0)$ and $\rm{sin}\theta(r\omega_0)$ are obtained without trigonometric functions for computational efficiency using the  relations
	\begin{eqnarray}\label{eqn:2.16}
	\rm{cos}\theta(r\omega_0) &=& \Re(Z(r\omega_0))/\abs{Z(r\omega_0)}, \\
	\rm{sin}\theta(r\omega_0) &=& \Im(Z(r\omega_0))/\abs{Z(r\omega_0)}
	\end{eqnarray}
   	Let $Y^{'}(r\omega_0)$ be the vector obtained after rotation of vector $Y(r\omega_0)$ by the phase difference $\theta(r\omega_0)$. Then the phase difference between $X(r\omega_0)$ and $Y^{'}(r\omega_0)$ reduces to zero or 
	\begin{eqnarray}\label{eqn:2.17}
	\phi_{X}(r\omega_0) - \phi_{Y^{'}}(r\omega_0) = 0
	\end{eqnarray}
	as required for a correctly calibrated system phase prior to formation of circular polarization.\\
\\2.4 Gain Equalization:         
The amplitudes also undergo linear distortions due to different gains in the two channels due to different passband characteristics. To compensate the amplitude differences, all the spectral components of $X_i(\omega)$ and $Y_i(\omega)$ in the passband are scaled to one same level, chosen to be the maximum signal level, $V_{max}$ in the passbands of $X_i(\omega)$  and $Y_i(\omega)$. Multiplying Eq.$(4)$ with its complex conjugate, we obtain the power spectrum $\abs{X_i(\omega)}^{2}$ and similarly Eq.$(5) $ yields $\abs{Y_i(\omega)}^{2}$. The scaling factors to equalize the magnitudes of $X_i(r\omega_0)$ and $Y_i(r\omega_0)$ are 	obtained from accumulating $\abs{X_i(r\omega_0)}^{2}$ and $\abs{Y_i(r\omega_0)}^{2}$ respectively for $T_{integ}$ to reduce thermal noise fluctuations in the measurement of the passband shapes and are expressed by the following two equations:
	\begin{eqnarray}\label{eqn:2.18}
	\abs{X(r\omega_0)}^{2} = \sum_{i=1}^{N_s}\abs{X_i(r\omega_0)}^{2}\\
	\abs{Y(r\omega_0)}^{2} = \sum_{i=1}^{N_s}\abs{Y_i(r\omega_0)}^{2}.
	\end{eqnarray}
	Expanding these using Eqs.$(6)$ and $(7)$ we find that each contains the following contributing terms for summation:
	\begin{eqnarray}\label{}
1.&\abs{H_X(r\omega_0)}^{2}&  \abs{S_i(r\omega_0)}^{2},\nonumber\\
&\abs{H_Y(r\omega_0)}^{2}&  \abs{S_i(r\omega_0)}^{2},\\
2.&\abs{H_X(r\omega_0)}^{2}&  \abs{N_{Xi}(r\omega_0)}^{2},\nonumber\\
&\abs{H_Y(r\omega_0)}^{2}&  \abs{N_{Yi}(r\omega_0)}^{2}
	\end{eqnarray}
where $\phi_{N_{Xi}}(r\omega_0)$ and $\phi_{N_{Yi}}(r\omega_0)$ are the phases of $N_{Xi}(r\omega_0)$ and $N_{Yi}(r\omega_0)$ respectively.\\
The first term (Eq.$17$) provides the measurement of the bandpass shape that we seek (assuming that the noise diode produces white noise). The second term (Eq.$18$) does not average to zero to the extent that noise sources other than the noise diode (eg receiver noise, radio sources, atmospheric emission, cosmic microwave background, and RFI) are present. We cancel this term, provided the corrupting source remains constant, by performing the summation in Eqs.$(15)$ and $(16)$ twice, first in the presence of $s_{i}(t)$ (noise diode switched on) and second with $s_{i}(t) = 0$ (noise diode switched off) and differencing. Hence, we are left with only the first term (Eq.$17$) where $\abs{H_X(r\omega_0)}^{2}$ and $\abs{H_Y(r\omega_0)}^{2}$ are the frequency-dependent amplitude scaling factors. To determine the gains $g_X(r\omega_0)$ and $g_Y(r\omega_0)$ that scale each $X_i(r\omega_0)$ and $Y_i(r\omega_0)$ to equalize the passband amplitudes to the same level, the following two equations are used
\begin{eqnarray}\label{} 
&g_X(r\omega_0)& = \sqrt{\frac{P_{max}}{\abs{X(r\omega_0)}^{2}}}, \\
&g_Y(r\omega_0)& = \sqrt{\frac{P_{max}}{\abs{Y(r\omega_0)}^{2}}}.
\end{eqnarray}
$P_{max}$ in Eqs.$(19)$ and $(20)$ is obtained by comparing all $\abs{X(r\omega_0)}^{2}$ and $\abs{Y(r\omega_0)}^{2}$ of Eqs.$(15)$ and $(16)$ for $r = 0,1,2,3,\ldots,N-1$ and finding the maximum power. The gains in Eqs.$(19)$ and $(20)$ are calculated using accumulated power spectra and are later applied to voltage spectra for equalization, which potentially introduces a small inaccuracy since the gains thus obtained are not the same as the actual gains obtained by accumulating individual voltage spectra consisting of absolute voltages in the denominators. However, if the absolute voltages, $\abs{X_j(r\omega_0)}$ and $\abs{Y_j(r\omega_0)}$, where $j= N_{s}+1,N_{s}+2,\ldots$ for the subsequent spectra acquired during observations, are accumulated for sufficient integration time then
\begin{eqnarray}\label{} 
g_X(r\omega_0) &\approx& \frac{V_{max}}{\sum\abs{X_j(r\omega_0)}} , \rm{and}\\
g_Y(r\omega_0) &\approx& \frac{V_{max}}{\sum\abs{Y_j(r\omega_0)}}.
\end{eqnarray}\\
$V_{max}$ in Eqs.$(21)$ and $(22)$ is the maximum of all $\sum\abs{X_j(r\omega_0)}$ and $\sum\abs{Y_j(r\omega_0)}$ for $r = 0,1,2,3,\ldots,N-1$. Eqs.$(21)$ and $(22)$ are also confirmed in numerical simulation.\\
\\
2.5 Windowing : 
A window function is derived for the $X_j$ and $Y_j$ spectra to truncate the possible analogue filter flanks to avoid scaling up, by large factors, signals that have been strongly attenuated by band-limiting filters. The window function is conveniently obtained from the $\abs{Z}$ spectrum since the $Z$ spectrum contains the band common to both $X$ and $Y$ spectra thereby providing necessary frequency shift and bandwidth information for the window function. Spectral channels in which the signal level $\abs{Z(r\omega_0)}$ is greater than one quarter of the maximum amplitude in the $\abs{Z}$ spectrum are given unit weight and all others are given zero weight resulting in a frequency shifted rectangular function of unit amplitude. A unit delta function is added to this since we want to pass undisturbed the DC signal produced by the A/D converter. This window function is applied to both $X_j$ and $Y_j$ spectra resulting in rectangular band shape for the frequency band that is in common. The effect of the window function on the time-domain is to convolve the  signal with a sinc function. Since, the frequency spectra are already band limited, the window function does not in itself introduce any new waveform characteristics. Rather it prevents contamination of the waveform that would arise were one to scale up, by large factors, frequency channels that had little signal.\\
\\
2.6 Forming Circular Polarization: 
The gains $g_X(r\omega_0)$ and $g_Y(r\omega_0)$, the rotation parameters $\rm{sin}\theta(r\omega_0)$ and $\rm{cos}\theta(r\omega_0)$, and the window function $W(r\omega_0)$ are applied to the spectral components $X_j(r\omega_0)$ and $Y_j(r\omega_0)$ respectively. If $Y_j^{''}(r\omega_0)$ and $X_j^{'}(r\omega_0)$ are the resulting vectors after calibration then $Y_j^{''}(r\omega_0)$ is related to $Y_j(r\omega_0)$ by the product of gain, window function and rotation matrix as follows
\begin{eqnarray}
\left[ \begin{array}{c} \Re(Y_{j}^{''}(r\omega_0)) \\ \Im(Y_{j}^{''}(r\omega_0)) \end{array} \right] &=& g_Y(r\omega_0)  W(r\omega_0)\left|\begin{array}{cc} \rm{cos}\theta(r\omega_0) & -\rm{sin}\theta(r\omega_0) \\ \rm{sin}\theta(r\omega_0) & \rm{cos}\theta(r\omega_0)  \end{array} \right|\nonumber\\
&\cdot& \left[ \begin{array}{c} \Re(Y_{j}(r\omega_0)) \\ \Im(Y_{j}(r\omega_0)) \end{array} \right]
\end{eqnarray}
	where $\Re(Y_{j}^{''}(r\omega_0))$, $\Im(Y_{j}^{''}(r\omega_0))$ and $\Re(Y_{j}(r\omega_0))$,  $\Im(Y_{j}(r\omega_0))$ are real and imaginary components of $Y_{j}^{''}(r\omega_0)$ and $Y_{j}(r\omega_0)$ respectively. Similarly,
\begin{eqnarray}
\left[ \begin{array}{c} \Re(X_{j}^{'}(r\omega_0)) \\ \Im(X_{j}^{'}(r\omega_0)) \end{array} \right] &=& g_X(r\omega_0)  W(r\omega_0)  \left[ \begin{array}{c} \Re(X_{j}(r\omega_0)) \\ \Im(X_{j}(r\omega_0)) \end{array} \right].
\end{eqnarray}

The windowed, phase and gain calibrated $X_j^{'}(r\omega_0)$ and $Y_j^{''}(r\omega_0)$ are added in quadrature ($\pm 90^{\circ}$) to obtain LHC and RHC polarizations.		   
\section{Performance Limitations}
In this section we will discuss the implications of possible contamination of phase caused by the cross polar component or leakage of unwanted orthogonal polarization component (D-terms) and by the temporal instability of receiver transfer characteristics and their effects on polarization purity with approximate quantitative results to estimate those errors. We will also discuss the requirement of frequent recalibration due to the variations in the transfer characteristics of the receiver by observing the drifts in the most sensitive parameters, which are the channel phases. Since the gain fluctuations are much smaller than phase fluctuations, we can ignore their effects.\\
\\ 
3.1 Effect of Phase Error on Polarization Purity:
Were one to introduce an imperfect $90^{\circ}$ phase shift into one channel when forming circular polarization from perfectly orthogonal linearly polarized channels, the output voltages $V_{RHC}$ and $V_{LHC}$ would contain unwanted contributions from the opposite hand of polarization.
\begin{eqnarray}
 &V_{LHC^{'}} =& V_{LHC} + (D_{LHC}  V_{RHC}) \\
&V_{RHC^{'}} =& V_{RHC} + (D_{RHC}  V_{LHC}) 
\end{eqnarray}\\
where $D_{LHC}$ and $D_{RHC}$ are the fractional voltage leakage factors from unwanted polarizations (D-terms).\\
The larger the phase error, the greater the contribution from the opposite hand. Consider the monochromatic case in which a linearly polarized wave is incident normally on crossed linear dipoles with the plane of the electric field oriented at $45^{\circ}$ to the two dipoles. Then the voltages in the two dipoles are
\begin{eqnarray}
 &V_x =& V_0 e^{j\omega t} \\
 &V_y =& V_0 e^{j\omega t}
\end{eqnarray}
 After introducing an imperfect $90^{\circ}$ phase shift to the $y$ channel, one has 
\begin{eqnarray}
&V_x =& V_0 e^{j\omega t} \\
&V_{y^{'}} =& \pm j V_0 e^{j\omega t} e^{j\epsilon}
\end{eqnarray}
where $\epsilon$ is the error in the $90^{\circ}$ phase shift. Circular polarization is obtained by summing the $x$ signal with the imperfectly phase-shifted $y$ signal, giving
\begin{eqnarray}
&V_{LHC^{'}} =& V_0 e^{j\omega t}(1-je^{j\epsilon}) \\
&V_{RHC^{'}} =& V_0 e^{j\omega t}(1+je^{j\epsilon}) 
\end{eqnarray}
Had the $90^{\circ}$ phase shift been perfect, then $\epsilon = 0^{\circ}$ giving $V_{LHC^{'}} = V_{LHC}$ and $V_{RHC^{'}} = V_{RHC}$.
Substituting Eq.($31$) and Eq.($32$) into Eqs.($25$) and ($26$) respectively, with $V_{LHC}$ and $V_{RHC}$ obtained by setting $\epsilon = 0$, we obtain the dependence of $D_{LHC}$ and $D_{RHC}$, on the phase error, $\epsilon$:
\begin{eqnarray}
D_{LHC} = -D_{RHC}= \frac{1 + \rm{sin}\epsilon - \rm{cos}\epsilon + j (1 - \rm{sin}\epsilon - \rm{cos}\epsilon)}{2} 
\end{eqnarray}
This result is used in section 3.3 for estimating the polarization purity.\\
\\
3.2 Phase Stability of the Analogue Receiver Chain:
The phase and amplitude transfer characteristics of the receiver
chains for the orthogonal polarization channels are known to drift
with time, due primarily to temperature changes of filters and cables
used in the receiver chains.  Fortunately, most of that change is
common to both orthogonal polarization channels as the equipment for
both channels is housed in close proximity to each other, and the
relative changes are small compared to the total.  The effect of drift
in the relative transfer characteristic is a degradation of the
polarization purity, since the equalizer weights that were determined
prior to an observation would no longer perfectly correct the channel
differences, by the amount of the relative drift since that
determination was made.  This translates into a requirement that the
equalizer weights be re-determined periodically to ensure that
polarization impurity due to drift remains below a pre-determined
level.\\
We have estimated how often such re-determination would need to be
made by measuring the relative phase drift in some existing receiver
chains at Effelsberg and the VLBA.\\
\begin{figure}[h]
\centering
\includegraphics[height=150mm, width=88mm]{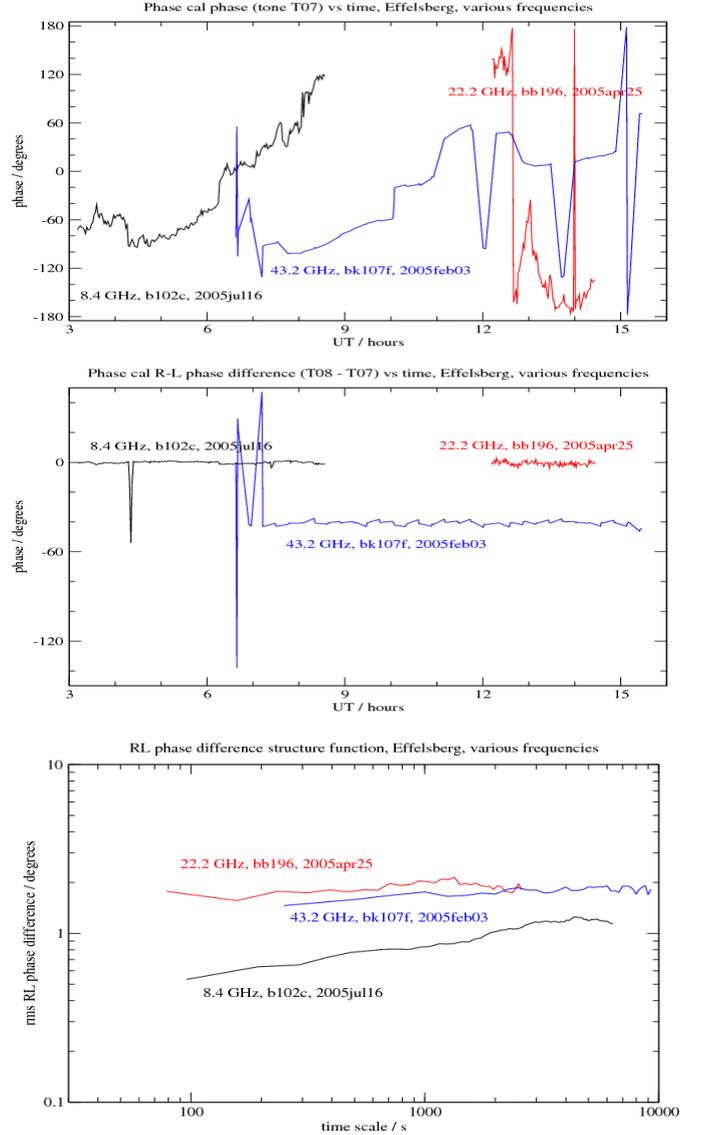}
\caption{{\it Top:} calibration system phase vs time for a single polarization
channel for three different receivers at Effelsberg, showing 
phase changes of typically $180^{\circ}$ to $360^{\circ}$  over 
periods of hours.
{\it Middle:} calibration system phase difference between orthogonal polarizations of
the same receivers at Effelsberg, for the same experiments as in the
top plot with an arbitrary offset. {\it Bottom:} structure functions constructed from the relative phases
presented in the middle plot.  These show the rms of the phase
difference vs time-scale.}
\end{figure}
The measurements were made using the VLBI phase calibration system \cite{thompson91},
which injects a pulse train in the front end and extracts them at the
backend data acquisition rack or correlator, to monitor the phase
length of the whole receiver system, from front end to the samplers.\\
The measurements show that indeed the phase changes in the
orthogonally-polarized channels track each other well (Fig 1 top and
middle) and there are only occasional outliers, most
likely related to phase-locked loop local oscillator used in the 
analogue base-band converters, which would not be present in a digital system.  The drift in the relative phase is conveniently quantified
using a structure function analysis, which converts the phase
difference time series into the rms phase change as a function of
time-scale (Fig 1 bottom).  The result is that the rms phase difference due
to drift is in the range $0.5^{\circ}$ to $2^{\circ}$ on time-scales of
$100$ s to $9000$ s.\\
\\3.3 Expected Polarization Purity:
The polarization purity to be expected from polarization conversion
performed at IF can be derived by combining the two results from section 3.1 and 3.2 -
the sensitivity of polarization leakage to phase errors and the typical phase errors in existing
analogue receiver chains ($0.5^{\circ}$ to $2^{\circ}$ rms).\\
The resulting D term is 0.006 (rms) for a $0.5^{\circ}$ rms phase error,
for which one must re-calibrate the equalizer every few minutes,
rising to 0.025 (rms) for a $2^{\circ}$ rms phase error, which one
would obtain were one to calibrate the equalizer once and leave it fixed for many hours.
These are smaller than the leakage D-terms measured for existing radio
telescopes, which are commonly 0.05 to 0.15.\\
However, the leakage in existing receivers is constant over long periods,
since it occurs primarily due to tolerances in the manufacture of the analogue
polarizers, and so can be calibrated using observations of astronomical 
polarization calibrators.  That calibration reduces the effect of the
leakage on the resulting polarization images by a factor of ten typically,
and one typically sees residual polarization artifacts that are 0.005 to 
0.015 rms times the peak flux density in the images.  These values are
comparable to those expected to be delivered from the digital polarization
conversion without use of astronomical D-term calibration.\\
However, the D-term from IF polarization conversion, though small,
is expected to drift with time between equalizer re-calibrations due
to drift in the relative phase of the orthogonal polarization receiver
channels.  Were one to want to improve on this by using astronomical 
calibration of the residual polarization leakage on time-scales between
the equalizer re-calibration, then one must be able to derive the D-term from a
snapshot observation.  Such an algorithm is available and requires
the use of an unpolarized calibrator source.  However, the changing
D-term requires that the post-correlation analysis software be able to
handle time-varying D-terms.\\
This will prevent the use of astronomical calibration of the
residual polarization leakage, since the D-term calibration in use assume that the 
leakage is constant on a 12 h time-scale.
\begin{figure*}
\centering
\includegraphics[height=120mm, width=170mm]{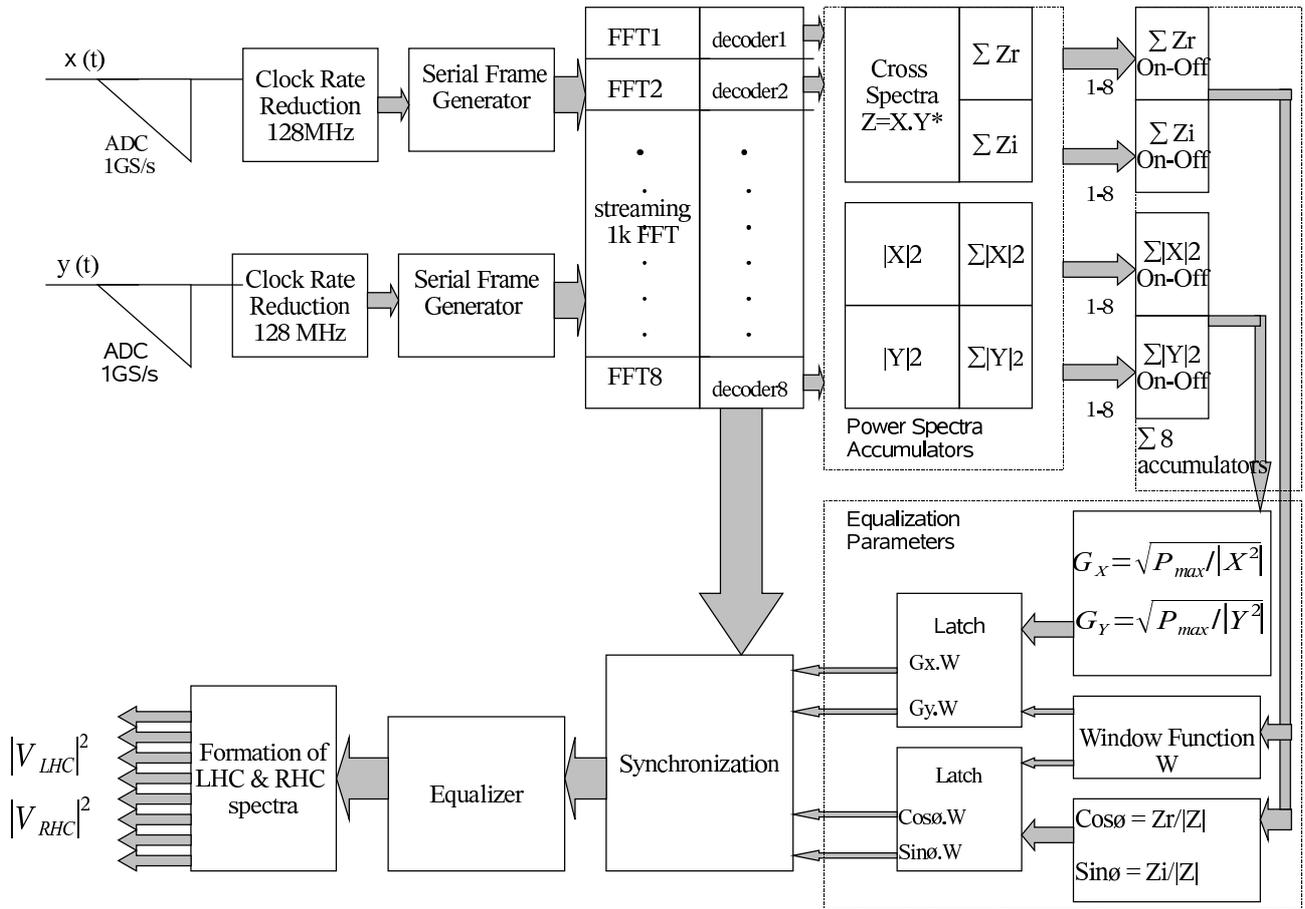}
\caption{Logic block for conversion from linear to circular polarization.}
\end{figure*}
\section{FPGA Implementation Details} 
Here we will describe the structural and functional details of the FPGA based circular polarizer (Fig. 2) for converting the two orthogonal input linear polarizations to the two circular polarization components in real time. The method of digital signal processing is based on the theory presented in section 2. The sampled time domain signals, $x$ and $y$ are fed, at a rate of $1024$ MS/s with 10 bits per sample representing positive integers, to the following logic elements.\\
\\
1. Clock rate reduction logic: Since the sampled time series at $1024$ MS/s is too fast for processing, serial-to-parallel conversion of the streaming data has been implemented with a factor-eight fanout, generating eight parallel sample streams clocked at $128$ MHz. This block was developed in an earlier DBBC (Digital Base-Band Converter) project. The front-end of the system operates with 1024 MHz sampling rate and the IF band is 512 MHz. The samples taken at 1024 MHz in the AD device are transferred in progressive steps in parallel, to reduce the clock rate, still maintaining a formal sampling clock of 1024 MHz. So the data are demultiplexed to obtain from 8-bit samples at 1024 MHz to 16-bit at 512MHz, 32-bit at 256 MHz, 64-bit at 128MHz (8 samples at 128 MHz). This system is already commercially available \cite{tuccari04}. 
 \\
\\2. Serial frame generator: This is central block for enabling serial processing of real time data in parallel working at 128 MHz clock rate. To feed the eight parallel streaming FFT blocks with frames of $1024$ real time-domain samples, intermediate logic takes in eight samples in parallel at every clock edge and outputs them to one of eight buffers, eight samples in parallel at each clock edge.  Once a complete frame of $1024$ samples have been loaded into the buffer, the next buffer is selected to receive the next $1024$ samples.  The first buffer is read out serially into its corresponding FFT block at one eighth of the rate ($128$ MHz) at which is was filled, starting at the next clock pulse after writing the first eight samples is complete, such that serial readout and parallel writing operations on the buffer occur simultaneously.  When the register is ready to read out the last sample of the first data frame, the first eight samples of frame number nine is ready for writing and the two operations occur simultaneously at the same clock pulse. This continues and each register sends out data serially at a rate of $128$ MHz.  The length of the frame is equal to twice the number of spectral channels required for later operations. \\
\\
3. FFT: The serial frame generator for the $x$ polarization and for the $y$ polarization each produce eight output lines to feed eight parallel FFT blocks that run continuously and independently of each other, each processing successive frames of data to keep up with the real-time sampling rate.  The FFT is a complex transform, but the data are purely real, and so we used the relation for the FFT of two real functions simultaneously, to save a factor of two in device resources by feeding the $x$ and $y$ polarization data to the real and imaginary channels of each FFT engine input.  The streaming pipelined FFT is generated conveniently using a Xilinx IP core. It is configured with 11 bits at the input and requires two's complement representation of the positive integers from the A/D converter. The real and imaginary terms of the $X$ and $Y$ spectra are retrieved from the FFT real and imaginary output channels by using a decoder at the output of each FFT.  The method is equivalent to performing separate FFTs of the two real functions, but consumes half the space by performing them simultaneously in one transform.  The FFT block outputs a $1024$-point complex spectrum every $10^{-6}$ s in each polarization, keeping up with the real-time $1024$ MS/s sample rate.\\
\\
4. Power spectra accumulator: To calibrate the equalizer weights, we used total-power spectrum accumulators when determining the amplitude scaling factors needed for bandpass calibration and we used cross-correlation and accumulation to determine the phase difference between the $X$ and $Y$ polarizations in each frequency channel.  To form the power spectra, each spectrum from each FFT (after decoding) is squared, forming $\abs{X}^{2}$ and $\abs{Y}^{2}$, and those are accumulated for nearly $8.4$ s, which is determined by the goal of having thermal noise fluctuations that contribute at most $0.1^{\circ}$ rms phase error during equalization.  The accumulator is dimensioned with enough bits to hold the accumulation result without overflow.  To measure the (frequency-dependent) phase difference between the $X$ and $Y$ polarizations, the cross-power spectra $Z = XY^{*}$  are formed and the real and imaginary terms, $Z_r$ and $Z_i$, are also accumulated for nearly $8.4$ s.  Each of the four quantities ($\abs{X}^{2}$, $\abs{Y}^{2}$, $Z_r$ and $Z_i$) are accumulated in eight pairs of accumulators, one following each FFT (after decoding), with each accumulator being paired to hold the noise diode on and off state results separately. After accumulation, the real-time processing of the calibration signal is stopped and calculations for the noise diode on and off states are performed sequentially on the accumulated results from all eight FFT engines. Differences are formed between the accumulators for the noise diode on and off states and the eight differences are summed to obtain the final integrated spectra, from which the equalizer gain and phase weights are calculated in the next block.\\
\\
5. Equalization parameters: The phase and gain equalization parameters are calculated using equations $(12), (13)$ and $(19), (20)$ respectively. The division and square root operations are implemented using the Xilinx floating point IP core.  The floating point operations are performed with a clock frequency of $64$ MHz due to the frequency limitations of the IP core. The resulting phase and gain calibration parameters are latched for real-time equalization of the $X$ and $Y$ data streams during subsequent observations.\\
\\
6. Window Function: A window function described in section (2.5) is determined and latched along with the equalization parameters.\\
\\
7. Synchronization: During observations the decoded FFT output samples are read out serially from the decoder and the corresponding gain, phase correction factors, and the window function are also read out serially from their respective latches and synchronization is required to ensure that the frequency channels of the spectrum and the equalization parameters are aligned. The synchronized values are passed to the next stage for phase and gain corrections.\\
\\
8. Equalization: Phase and gain corrections are applied to each spectral channel of the outputs after synchronization in the equalizer block, which implements equations ($23$) and ($24$).\\
\\
9. Formation of circular polarization:   After equalization $X$ and $\pm 90 ^{\circ}$ phase shifted $Y$ are added to form the LHC and RHC polarizations.  The $\pm90^{\circ}$ phase shifts of $Y$ are implemented by exchanging the real and imaginary components of each spectral channel in the $Y$ spectrum with a sign inversion of the real component for $+90 ^{\circ}$ phase shift and  of the imaginary component for $-90^{\circ}$ phase shift. The number of bits of each spectral channel in each output power spectrum is 51.\\
\\
The above logic blocks are tested separately since the design is too big to fit one FPGA chip. The design (Fig. 2) is also broken in parts to fit the commercially available DBBC boards and a practical implementation will be carried out in another future project by stacking DBBC boards each containing a part of the broken design. The simplest configuration requires fourteen such boards to keep the required precision and since DBBC has defined number of input/output pins. Since the DBBC already converts 1GS/s to 8 samples at 128 MHz and the logic blocks are tested to work at 128 MHz, we are certain that the developed circular polarizer will show expected performance in real time. The cost of implementation in the DBBC is around 50,000 euros and in custom designed boards is around 30,000 euros.
\section{Verification of Polarization Purity}
We performed the following experiment to simulate the developed circular polarizer to quantify the polarization purity obtained by this digital technique. The experiment was performed in an anechoic chamber by coupling linearly polarized broad-band noise into a circular waveguide and receiving with crossed linear dipoles at the end of the waveguide. The calibration coupler, which is responsible for polarization purity was tuned for equal amplitude and phase. The cross coupling was estimated to be better than -33.6 dB while doing the setup. The experimental setup is shown in Fig. $3$.
\begin{figure}[h]
\centering
\includegraphics[height=70mm, width=90mm]{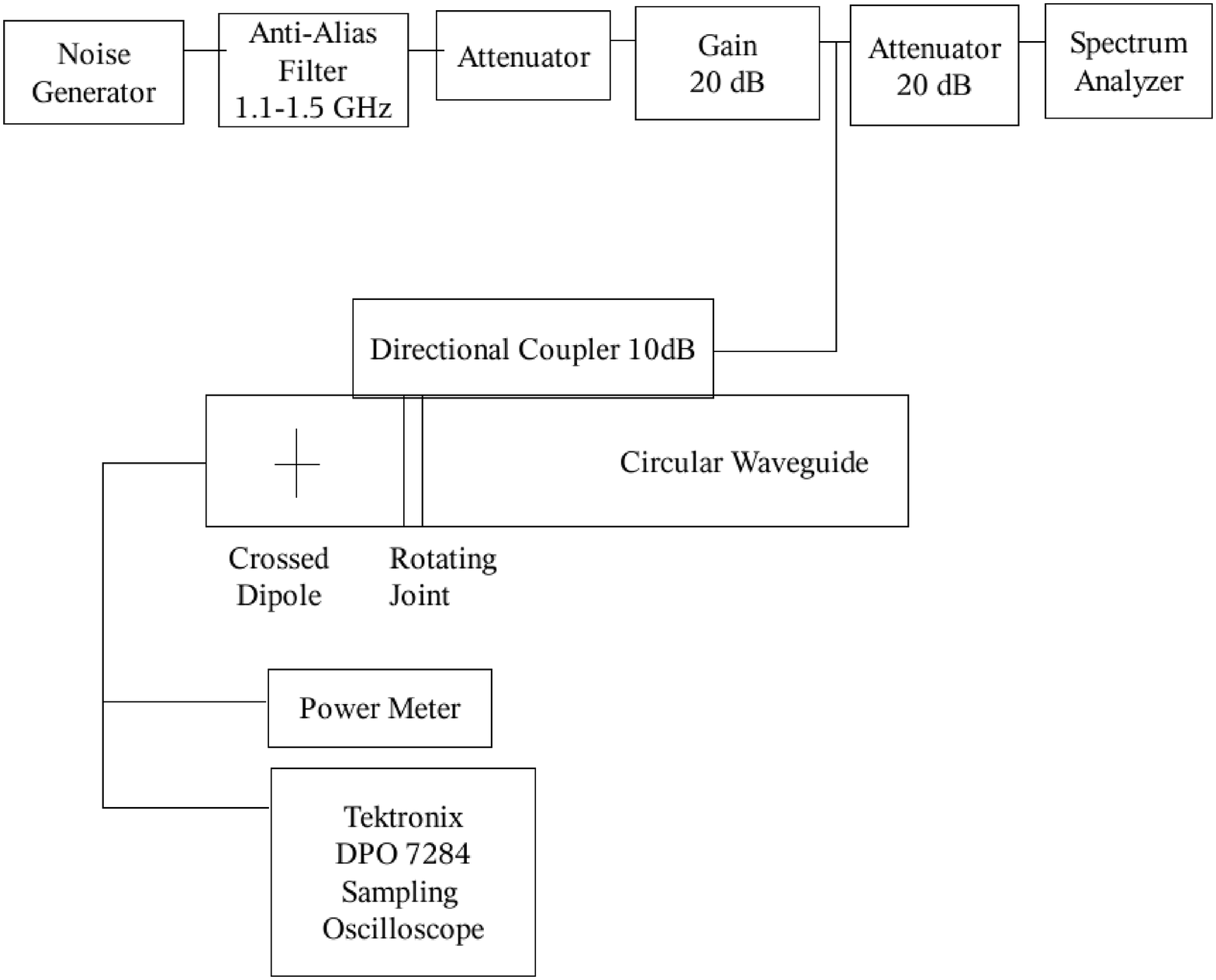}
\caption{Experimental setup in anechoic chamber to measure polarization purity.}
\end{figure}
\begin{figure}[h]
\includegraphics[height=70mm, width=90mm]{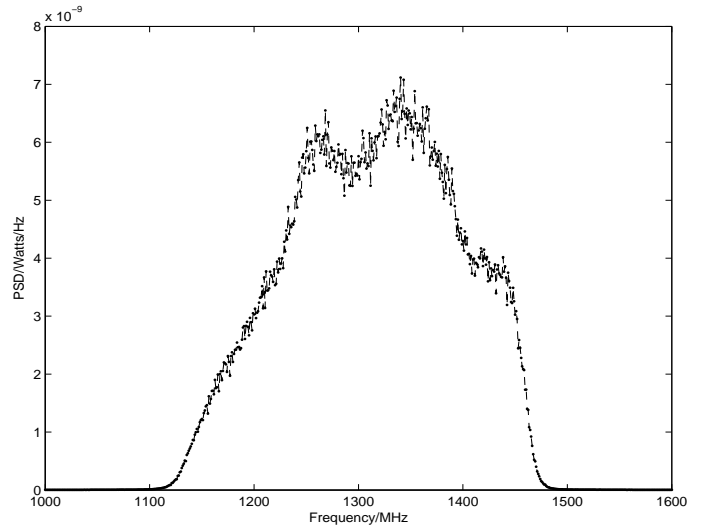}
\caption{Power spectrum of the broad-band noise measured at the input to the directional coupler by the spectrum analyzer shown in Fig. 3.  The band was shaped using an anti-aliasing filter to ensure that the power level was low below $1000$ MHz and above $1500$ MHz to avoid aliasing of power from outside the third Nyquist zone during the later digital down conversion.}
\end{figure}
The spectrum of the broad-band noise that was coupled into the waveguide is shown in Fig. 4.  The noise power spanned from $1160$ MHz to $1462$ MHz ($6$ dB down from peak), representing a fractional bandwidth of $23$ \%. The power level at the edges of the Nyquist band, at $1000$ MHz and $1500$ MHz, were $37$ dB and $30$ dB below the peak in the band, and so very little power was aliased from outside the band. After receiving this signal with the crossed dipoles, the signal was under-sampled at a sample rate of $1000$ MS/s by a digital oscilloscope, causing digital down conversion to baseband. Since the sample rate was slightly less than 1024 MS/s used by the design, the frequency channel width in this experiment was 0.976 MHz instead of 1 MHz. The input voltage range of -250 mV to +250 mV was translated to 0 to 1024 representing 10 bits positive integers before feeding the design for digital processing. No special effort was taken to match the complex gain or path length in the two channels from the receiving dipoles to the sampling oscilloscope. Thus, the RF band was translated to $160$ MHz to $462$ MHz, representing a fractional bandwidth of $97$  \% that was presented to the polarization converter.\\
\begin{figure}[h]
\centering
\includegraphics[height=150mm, width=90mm]{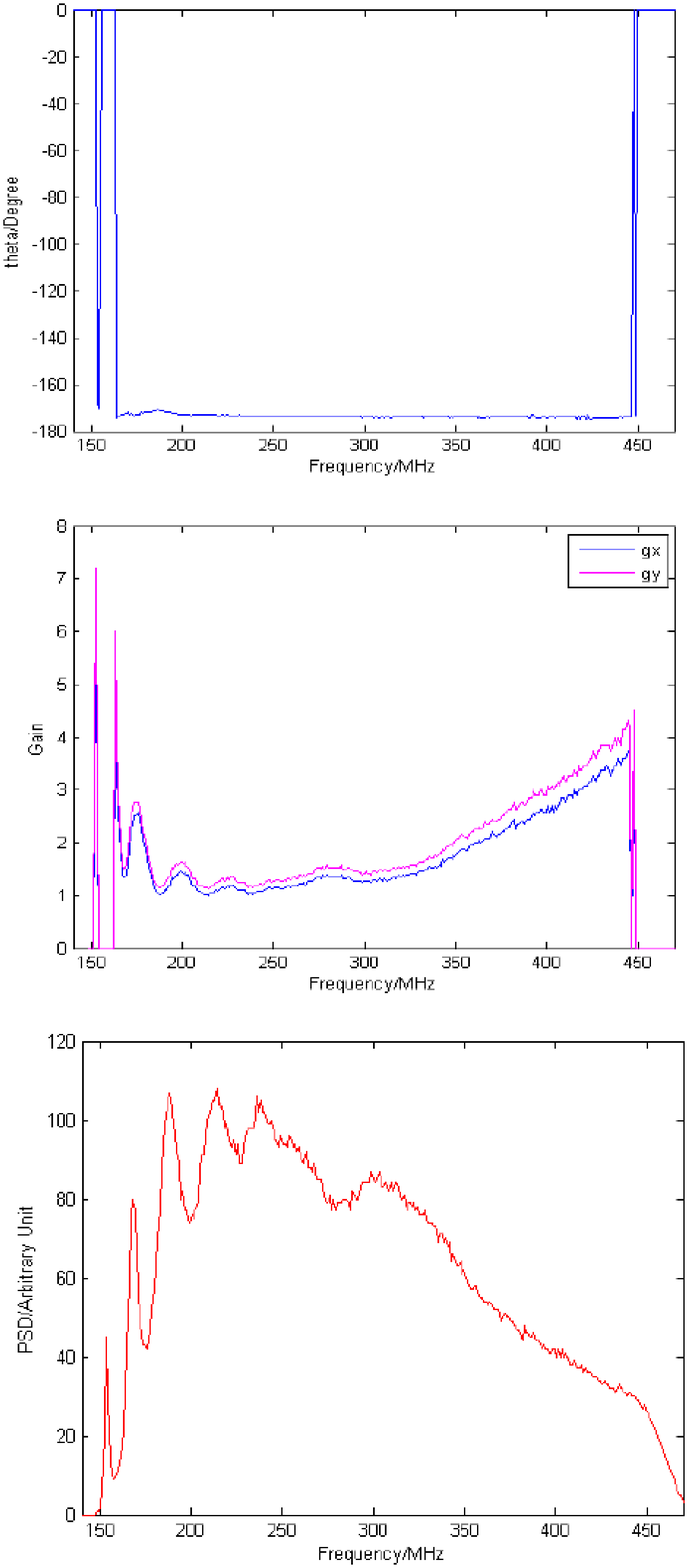}
\caption{Equalizer weights determined during calibration of the equalizer on the noise diode injected at $45^{\circ}$ to the two receiving dipoles. {\it Top:} phase difference between {\it X} and {\it Y} polarization channels. {\it Middle:} gains for {\it X} and {\it Y} polarization channels.  {\it Bottom:} power spectrum of the broad-band noise measured at the dipole outputs by the oscilloscope. The frequency range is labeled corresponding to the band after digital down conversion from the RF band of (1000 to 1500) MHz to baseband of (0 to 500) MHz.  The gains are approximately the inverse of the square root of the noise spectrum in the bottom as expected. The gains and phases are set to zero by the window function where the power dropped 6 dB below the maximum, which happened below 150 MHz and above 450 MHz.  This is as expected, given the shape of the broad-band noise spectrum.}
\end{figure}
1. Measurement Description:
The outputs of the crossed linear receiving dipoles were sampled with a digital storage oscilloscope for $4$ ms and saved to file for later processing by the design.  A rotating waveguide joint allowed us to rotate the plane of linear polarization with respect to the crossed receiving dipoles. Any ellipticity of the resulting circular polarization would show up as a change in the power in the circular polarization as the linear polarization is rotated (see section $6.2$).\\
\\
2.  Equalizer Calibration:
For equalizer calibration, the linear polarization was aligned at $45^{\circ}$ to the two dipoles by connecting the dipole outputs to a two-channel power meter and rotated until equal power was measured in both channels. The resulting powers were  $+7.51$ dBm 	and $+7.50$ dBm and drifted by $\pm 0.14$ dB during the measurement, corresponding to a polarization rotational position angle uncertainty of $0.9^{\circ}$. For the noise diode off state, the data samples were simply set to zero rather than being measured, since the setup in Fig. 3 was well shielded from RFI, in which case measuring with the noise diode off gives almost same results as setting the off-state samples to zero (verified in a previous experiment). The signals were sampled by the digital oscilloscope with the noise diode on and processed by the design to obtain the equalizer amplitude and phase weights.  These were loaded into the equalizer for  calibration of subsequent measurements, and are shown in Fig. $5$. The power spectrum in Fig. $5$ (bottom) differs from the spectrum in Fig. 4 since the transfer characteristics of the sampled spectrum are modified by the frequency response of each component through which the signal is transferred namely the directional coupler, circular waveguide, dipoles and the sampling oscilloscope.\\
\\
3. Polarization Purity Measurement:
For polarization purity measurement, the plane of input linear polarization was rotated to five positions with respect to the receiving dipoles and the received signals from both dipoles were sampled in each position.  To adjust the rotational position angle accurately, the plane of input linear polarization was rotated to either minimize the power in one of the receiving dipoles ($90^{\circ}, 0^{\circ}, -90^{\circ}$) or to obtain equal power in both dipoles ($45^{\circ}, -45^{\circ}$). The time elapsed between calibrating the equalizer and making all the measurements for determining the polarization purity was some 2 h.  During this time, some drift in components might have occurred, but nevertheless good polarization purity was obtained.\\
\\
4. Data Processing to Form Circular Polarization: 
We processed the data by running the design in software logic simulation, using ModelSim SE on suse 10.3 Linux machines having 16 GB of RAM and 2.7 GHz clock speed. Processing of $4$ ms of data from each position required $4$ days of elapsed time on a single computer so we processed for each position 20 sets of $1/20^{th}$ of the data in parallel, which required one day.
\section{Results}
The resulting power spectra in LHC and RHC are shown in Fig. $6$ for one of the five position angles of the input linear polarization on top of each other.  These show that the gain equalization flattened the spectra and that the window function truncated the spectra where the filters roll off. The total power is found to change very little with rotation of the input linear polarization, signaling a high purity circular polarization.  To quantify the purity, we measured the total power in LHC and RHC as a function of rotation angle, and show this in Fig $7$.  This shows power level changes of around $1$ part in $200$ peak-to-peak over all position angles as a function of rotation angle. We obtained an ellipticity of 0.9971 and 0.9976, axial ratio of 1.0029 and 1.0024 cross-polar response of -24.975 dB and -24.979 dB and D-term of 0.05767 and .05764 for LHC and RHC polarizations respectively. Since the cross coupling of -33.6 dB caused by the directional coupler is better than the obtained cross-polar response, the limiting factor is most likely the repeatability of the connections between matings. The insertion loss is unlikely to be repeatable between matings to -34 dB that would cause the calibration signal to be the limiting factor. The uncertainty on the polarization purity measurement has contributions from thermal noise, mechanical tolerance in the rotating waveguide joint, and in the repeatability of RF connections that we had to disconnect and reconnect during the measurement for connecting alternately the power meter and the oscilloscope.  These can be estimated from the small changes of the power in the circular polarization formed when the input linear polarization was at $-90^{\circ}$ and at $90^{\circ}$. Those two positions are symmetric and the resulting powers should be equal, regardless of the ellipticity of the transmitting or the receiving antenna.  We found fractional changes of 0.001 in the LHC power and 0.0004 in the RHC power between these two position angles.  These are comparable to the peak power variations seen as we rotated the input linear polarization, and so the polarization leakage measurement is limited by mechanical tolerances in the apparatus.  The thermal noise contribution was minor - the fractional error due to thermal noise fluctuations in the total power measurement was only 0.0006.
\begin{figure}[h]
\centering
\includegraphics[height=60mm, width=90mm]{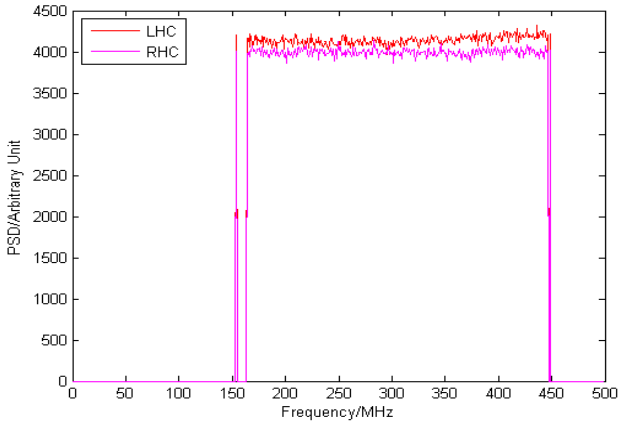}
\caption{Power spectral densities for LHC and RHC for a single position angle.}
\end{figure}
\\Discussion:
A small systematic offset is seen between LHC and RHC powers (Fig. 6). A small error in the magnitudes of the rotation matrix elements will cause the phase rotation of $Y$ to introduce an offset between the LHC and the RHC powers. This can happen since truncation error can cause small differences in the obtained phase difference and the  actual phase difference. The following equations show the dependence of the offset on the rotation angle error 
$\theta_{\epsilon}(r\omega_0) =\theta_Y(r\omega_0) + \theta(r\omega_0) - \theta_X(r\omega_0)$.\\
\begin{eqnarray}
V_{LHC}(r\omega_0) &=& X'(r\omega_0) - jY''(r\omega_0)\\
	&=& g_{X}(r\omega_0)\abs{X(r\omega_0)}e^{j\theta_X(r\omega_0)}\nonumber\\
	&-& j g_{Y}(r\omega_0)\abs{Y(r\omega_0)}e^{j\theta(r\omega_0)}e^{j\theta_Y(r\omega_0)}\\
	&=& V_{max}e^{j\theta_X(r\omega_0)}(1-je^{j\theta_{\epsilon}(r\omega_0)})
\end{eqnarray}
Therefore, 
\begin{eqnarray}
\abs{V_{LHC}(r\omega_0)}^2 = V_{max}^2(2 + 2sin\theta_{\epsilon}(r\omega_0))
\end{eqnarray}  \rm{Similarly,}
\begin{eqnarray}
V_{RHC}(r\omega_0) &=& X'(r\omega_0) + jY''(r\omega_0) \\
	&=& V_{max}e^{j\theta_X(r\omega_0)}(1+je^{j\theta_{\epsilon}(r\omega_0)})\\
\abs{V_{RHC}(r\omega_0)}^2 &=& V_{max}^2(2 - 2sin\theta_{\epsilon}(r\omega_0))
\end{eqnarray}
Hence, the offset between the two power spectra is given by
\begin{eqnarray}
\abs{V_{LHC}(r\omega_0)}^2 - \abs{V_{RHC}(r\omega_0)}^2 = 4V_{max}^2sin\theta_{\epsilon}(r\omega_0)
\end{eqnarray}
for a single channel.\\
\begin{figure}[h]
\centering
\includegraphics[height=60mm, width=90mm]{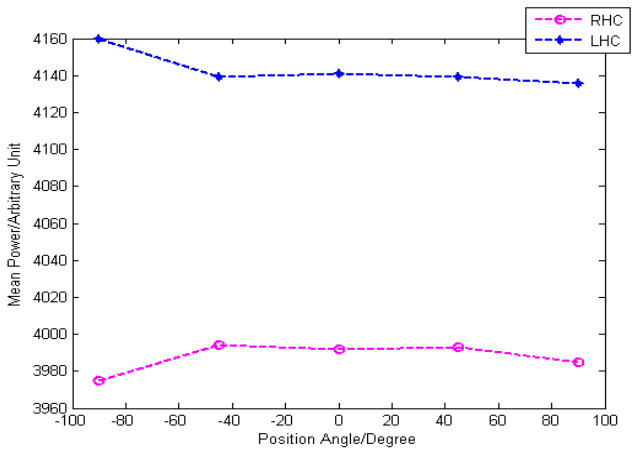}
\caption{Mean output powers in LHC and RHC as a function of the rotational position angle of the input linear polarization.}
\end{figure}
\\The RHC power is almost flat across band but LHC showed a small change with frequency.
To quantify this variation, we took equally spaced points from each power spectrum in the band of interest and plotted the deviation of those points from the mean of the respective power spectrum. Fig. 8 shows the deviation of LHC and RHC power spectra from their mean values. The plots are derived from LHC and RHC power spectra obtained by applying the equalization parameters to the same data samples from which the equalization parameters are obtained to show the deviations caused by bit truncation. From around 300 MHz onwards (Fig. 8 bottom) the two polarizations deviate differently with increasing frequency. This difference can be explained in terms of truncation of numerical precision. We have truncated bits at stages before and after formation of LHC and RHC powers, determined by the input signal levels and by the aim to keep phase errors $< 0.1^{\circ}$. The truncation error will cause offsets in the two circular power spectra shown in Eqs. (37) and (40), to deviate unequally from their mean and hence, the deviation can be more in one than the other. We verified in simulations that the offsets and the distortions seen in LHC and RHC power spectra increases in proportion to the number of bits discarded. However, the effects are small and we nevertheless obtained good polarization purity.
\\
\begin{figure}[h]
\centering
\includegraphics[height=100mm, width=90mm]{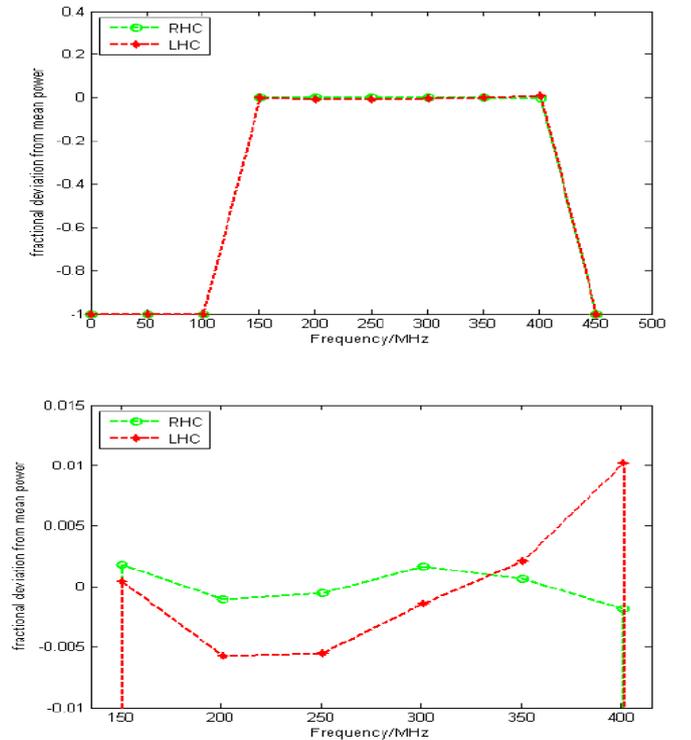}
\caption{{\it Top:} fractional deviations of LHC and RHC powers from their mean values across the band of interest. The spikes at 160 MHz in the two spectra (Fig. 6) are excluded. {\it Bottom:} same as top, but zoomed in to show the trends of the deviations in the two power spectra.}
\end{figure}
\\
The equalizer applies phase correction on 1 MHz channel spacing, so we must ensure that when the VLBI correlator later subdivides the spectrum into finer frequency channels that the phase equalization is smoothly interpolated and does not result in a sawtooth with 1 MHz spacing.  We performed a numerical simulation that confirmed that when we applied equalizer weights at 1 MHz spacings, transformed back to time domain, then transformed to frequency domain with 0.5 MHz spacing (by adjoining two 1024 point frames of time domain data) then the intermediate points at $n + 0.5$ MHz, where $n$ is an integer were seen also to have been phase corrected by the equalizer phase.
\section{Conclusion}
Though this technique was demonstrated using logic simulator in software, it has been 
implemented using Xilinx software generating firmware that can be loaded
into FPGAs.  An implementation on the digital baseband converter \cite{tuccari08} is in preparation,
which will make this technique available for use at many radio observatories for VLBI, and with minor
extension, for measuring Stokes parameters.
A further extension offers the prospect of cancelling the remaining (already low) cross polarization
response (D-term).  The idea is that one can use adaptive cancellation by forming a linear combination
of the two polarizations and adapting the coefficients of the filter to minimize the output power.
The result should be (almost) perfect polarization purity in circular polarization over large fractional
bandwidths.  This could enable the sensitive search for circular polarization in active galactic nuclei.\\
Later measurements at the telescope should confirm that excellent polarization purity is achieved in real applications, as it was in the anechoic chamber.  One can also confirm the stability of the transfer characteristics and decide on a re-calibration interval for operational use.  One can also characterize the typical phase response of receivers and confirm that the choice of 1 MHz frequency spacing is well matched to the existing systems. The effect of RFI on the system can be explored, to give recommendations on tolerable RFI levels and required performance of mitigation strategies. The trend in next-generation receivers for radio astronomy is to move the samplers as close as possible to the front end, which will benefit this system of polarization conversion since the time variable path length changes due to analogue cables and filters and amplifiers will be much reduced, yielding even better polarization purity.\\
\\
\rm{Acknowledgements}  \rm{We gratefully acknowledge Xilinx for their generous support, through their donation of 
FPGAs for this project and for providing prompt help with software issues as they arose.\\
Koyel Das, as a member of the International Max Planck Research School for Astronomy and Astrophysics at the Universities of Bonn and Cologne, thankfully acknowledges IMPRS for funding her PhD projects}

\end{document}